# Radiation Produced with Slow-Wave Fundamental Mode and Generalized Fundamental mode in Periodic Structures

Yifan Yin[1,2,3], Shunli Li[3], *Member IEEE* and Ke Wu[2], *Fellow, IEEE*

*Abstract*—This study suggests an idea that radiation in a periodic leaky-wave antenna (PLWA) should be considered to be produced with the fundamental mode, regardless of whether it is fast-wave or slow-wave. The idea is different from the conventional PLWA theory, which considers it a fact that a PLWA produces radiation with its fast-wave space harmonic when the fundamental mode is slow-wave. To elaborate the idea, it is proved that there is not an eigen-equation like Pythagorean theorem for the fundamental mode in PLWAs. Then a non-uniform structure antenna is designed to show that slow-wave modes can produce leaky-wave radiation. Again it is proved that the difference of the phase constants between a slow-wave fundamental mode and its fast-wave space harmonics has not any effect on the radiation pattern of a PLWA. Moreover, it is clarified that the fundamental mode has a more definite physical significance than space harmonics. Finally, a concept of generalized fundamental modes is proposed without using Fourier expansion. The generalized fundamental modes have the same phase and attenuation constants as any space harmonics, and have physical significance as the conventional fundamental mode. Therefore, it could replace the roles that the space harmonics used to play.

## I. INTRODUCTION

LEAKY-WAVE radiation is well known to takes place in both uniform and non-uniform leaky-wave antennas (LWAs) [1]. A uniform LWA, i.e. a rectangular air-filled waveguide with a longitudinal slot on its broadside wall, is a uniform transmission structure in which there are a dominant mode or high-order modes [2]-[3]. As a general perception, a uniform LWA produces leaky-wave radiation with its fundamental or high-order modes, which must be fast-wave [1]-[7]. What about a non-uniform LWA is？

In general, there are two types of non-uniform LWAs, namely periodic leaky-wave antennas (PLWAs) [8]-[15] and non-periodic leaky-wave antennas (NPLWAs) [16]-[17].

A PLWA is no longer a uniform but a periodic structure, in which there are a fundamental mode and space harmonics [4]-[5],[18]. When the fundamental mode is a fast-wave, a PLWA produces fast-wave radiation with the fundamental mode. When the fundamental mode is slow-wave, in the classic theory of PLWA, it has always been thought of the fact that the PLWA produces slow-wave radiation with the space harmonic which should be fast-wave [1], [4]-[5], [8]-[10], [18]-[24]. It is natural to ask why the slow-wave radiation of a PLWA cannot be produced with the slow-wave fundamental mode, but with fast-wave harmonics? Is it because fast-wave modes cannot produce radiation?

For the mode in a uniform LWA, there is an eigen-equation like Pythagorean theorem, which governs the three dimensional wave-numbers of the mode. Based on the eigen-equation, it can be concluded that a fast-wave mode cannot produce radiation im a uniform LWA. Is there such a eigen-equation for the fundamental mode in a PLWA, with which the slow-wave the fundamental mode would not produce slow-wave radiation. However, to the best knowledge of the authors, no one has yet proved the eigen-equation for the fundamental mode.

On the other hand, there is work demonstrating that slow-wave dominant modes in non-uniform LWAs can produce radiation [16]-[17]. The antennas in [16] and [17] actually are two NPLWAs, which means that there is no space harmonics at all. The dominant modes in these antennas are slow-wave, so it is reasonable to think that the slow-wave dominant modes produce radiation of these antennas.

This paper suggests an idea that the radiation of a PLWA should be considered to be produced with the fundamental mode, regardless of whether it is fast-wave or slow-wave [25]. The idea is different from the conventional PLWA theory, and the three prerequisites of the idea are demonstrated as follows:

1) Slow waves in a non-uniform structure antenna could produce radiation. To verify prerequisite, it is proved in Section II that there is not an eigen-equation like Pythagorean theorem for the fundamental mode, and a non-uniform antenna is

Yifan Yin is with the Jiangsu Key Laboratory of Wireless Communications, Nanjing University of Posts and Telecommunications, Nanjing 210003, China, with the Department of Electrical Engineering, Poly-Grames Research Center, Polytechnique Montreal, Montreal, QC H3T 1J3, Canada, and also with the State Key Laboratory of Millimeter Waves, Southeast University, Nanjing 210096, China (e-mail: yinyf@njupt.edu.cn).
Shunli Li is with the State Key Laboratory of Millimeter Waves, Southeast University, Nanjing 210096, China (e-mail: 101012159@seu.edu.cn).
Ke Wu are with the Department of Electrical Engineering, Poly-Grames Research Center, Polytechnique Montreal, Montreal, QC H3T 1J3, Canada (e-mail: ke.wu@polymtl.ca).

designed in Section III to show that its slow-wave dominant mode can produce leaky-wave radiation.

2) It is proved in Section IV that the difference of the phase constants between a slow-wave fundamental mode and its fast-wave space harmonics has not any effect on the radiation pattern of a PLWA.

3) The fundamental mode has a more definite physical significance than space harmonics. This prerequisite is elaborated in Section V.

The paper also proposes a concept of the generalized fundamental modes without using Fourier expansion in Section VI. The generalized fundamental modes and the fundamental mode differ by only a phase factor. Since the generalized fundamental modes can have the same phase constant and attenuation constant as any space harmonic, it is suggested that the generalized fundamental modes could replace the roles that were once played by the space harmonics

In Section VII a conclusion is provided.

## II. IS THERE EIGEN-EQUATION FOR FUNDAMENTAL MODE IN A PERIODIC STRUCTURE

This Section will proof that there is not an eigen-equation like Pythagorean theorem for the fundamental mode. As a result, whether the fundamental mode radiates or not, it cannot be determined based on whether the fundamental mode is fast-wave or slow-wave.

Appling Floquet's theorem to a periodic structure like a PLWA, one can write the field in the periodic structure as follows [18]

$$\mathbf{E}(x, y, z) = \mathbf{E}_0(x, y, z)e^{-j\beta_0 z} \quad (1)$$

$$\mathbf{E}_0(x, y, z + p) = \mathbf{E}_0(x, y, z) \quad (2)$$

where the function of the fundamental mode, $\mathbf{E}_0(x, y, z)$, is a periodic function with a periodicity $p$ about the argument $z$, $\beta_0$ is the propagation constant of the fundamental mode.

Substituting (1) into wave equation, a three-dimensional wave equation for the fundamental mode is as follow

$$(\nabla_T^2 + \frac{\partial^2}{\partial z^2} - \beta_0^2 + k_0^2)\mathbf{E}_0(x, y, z) - 2j\beta_0 \frac{\partial \mathbf{E}_0(x, y, z)}{\partial z} = 0 \quad (3)$$

Rather than a two-dimensional partial equation that the eigenmode in a uniform transmission structure satisfies, the equation (3) generally is a three-dimensional partial differential equation. The propagation constant $\beta_0$ depends on the boundary condition on the whole boundary of a unit cell rather than the transverse boundary condition only on its transverse section.

Only if the following equation

$$(\frac{\partial^2}{\partial z^2} - 2j\beta_0 \frac{\partial}{\partial z})\mathbf{E}_0(x, y, z) = 0 \quad (4),$$

holds, the equation (3) would become a two-dimensional partial equation. The solution to the equation (16) is as follow

$$\mathbf{E}_0(x, y, z) = j\frac{\mathbf{E}_{01}(x, y)}{2\beta_0} + \mathbf{E}_{02}(x, y)e^{j2\beta_0 z} \quad (5).$$

Substituting (5) into (1), one can write the field as follow

$$\mathbf{E}(x, y, z) = j\frac{\mathbf{E}_{01}(x, y)}{2\beta_0}e^{-j\beta_0 z} + \mathbf{E}_{02}(x, y)e^{j\beta_0 z} \quad (6).$$

The field in (6) actually is the electromagnetic modes in a uniform rather than periodic transmission structure. Therefore, for the fundamental mode in a periodic structure there is an inequation as follow.

$$(\nabla_T^2 - \beta_0^2 + k_0^2)\mathbf{E}_0(x, y, z) \neq 0 \quad (7).$$

Consequently, the $\mathbf{E}_0(x, y, z)$ cannot be characterized with only a pair of transverse eigen parameters of $k_x$ and $k_y$, which are usually determined only by the transverse boundary condition on its transverse section in a uniform structure. In fact, because a periodic structure always has variable cross-sections along its propagation direction, one can not obtain a fixed propagation constant $\beta_0$ governing the whole periodic structure, because there is not a constant transverse section to enforce boundary conditions. Therefore, there is certainly an inequation for the fundamental mode as follow

$$k_0^2 \neq k_x^2 + k_y^2 + \beta_0^2 \quad (8).$$

If a fundamental mode is a slow-wave, namely $\beta_0 > k_0$, due to inequation (8), the $k_x^2 + k_y^2$ cannot be determined to be negative, and the $k_x^2 + k_y^2$ might be positive! When the real part of the $k_x$ or/and $k_y$ is not zero, the fundamental mode has phase variation along its transverse direction, and can produce leaky-wave radiation. As a result, the slow-wave fundamental mode could be responsible for leaky-wave radiation.

## III. CAN SLOW-WAVE PRODUCE RADIATION IN NON-UNIFORM LWA

To demonstrate that the slow-wave dominant mode in a non-uniform structure antenna can produce radiation, this Section introduces a very simple traveling-wave series-fed array antenna, namely slotted dielectric filled waveguide (DFW) antenna. Fig. 1(a) shows the geometries structures of the slotted DFW antenna, and Fig. 1(b) is the photo of its SIW prototype. The SIW has a equivalent width of 17.9 mm. and is designed on a Rogers RO4003 dielectric substrate ($\varepsilon_r = 3.55$, $\tan \delta = 0.0022$). Both the DFW and SIW here are 75.4 mm longs. Since the slotted DFW/SIW has only two slots, it cannot be a periodic structure and there is no any space harmonics in it.

To simplify the following simulations with CST, without otherwise specified, all the substrates are treated as lossless and the metal is treated as PEC (perfect electric conductor).

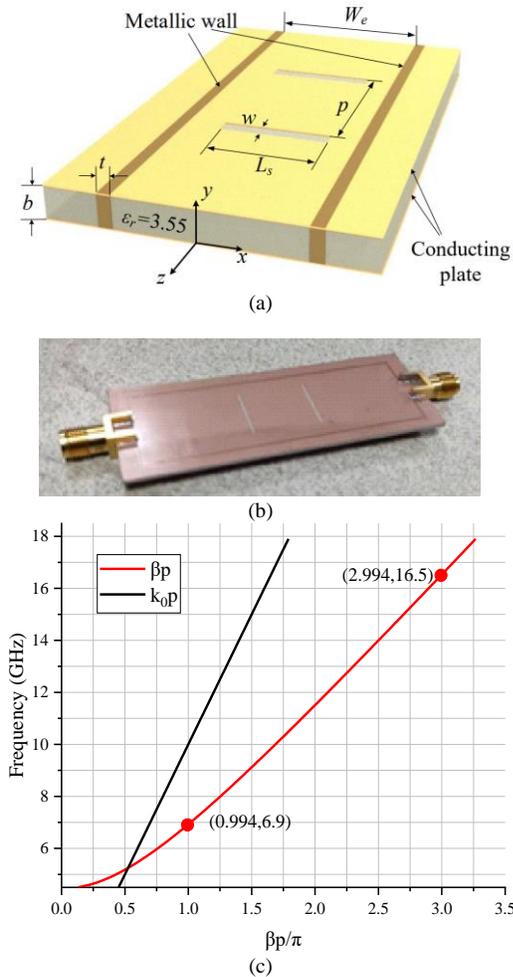

Fig. 1. Slotted DFW/SIW: (a) Structure, (b) Its SIW prototype and (c) Dispersion (Ws=18 mm, s=0.3 mm, d=0.6 mm, We=17.9 mm, b=0.813 mm, w=0.4 mm, Ls=12 mm, p=15 mm. and $t$=0.6 mm).

When the dominant mode $TE_{10}$ is fed into a DFW, its phase constant $\beta$ can be calculated as follows [18],

$$\beta = k_0\sqrt{\varepsilon_r - \left(\frac{\lambda_0}{2W_e}\right)^2} \quad (9)$$

where $\lambda_0$ and $k_0$ are wavelength and wavenumber in air, respectively. Fig. 1(c) shows the dispersion of the DFW, calculated by equation (9). It indicates that the phase delay, $\beta p$, between the two slots is $0.994\pi$ and $2.994\pi$ at 6.9 GHz and 16.5 GHz, respectively. In addition, when the operating frequency exceeds 5.4 GHz, the dominant mode is slow-wave.

The two-slot DFW can be treated as a two-element array. The radiation pattern of the two-element array can be written as [26]

$$F_\theta = F_0(\theta)e^{-j(kr+0.5\beta p)}\cos[0.5p(k_0\cos\theta - \beta)] \quad (10)$$

where $F_0(\cdot)$ is the radiation pattern of a single slot. Based (10), it can be find that when $\beta p$ is $\pi$ and $3\pi$, the radiation at broadside is null.

Fig. 2(a) and (b) show the simulated 3D radiation patterns of the DFW, and Fig. 2(c) and (d) plot the simulated and measured radiation patterns on E-plane of the two-slot SIW. All these patterns have a null radiation at broadside at 6.9 GHz and 16.5 GHz, respectively.

Fig. 2(e) and (f) show the simulated E-field distribution on plane $y=0.5b$ inside the two-slot DFW at 6.9 GHz and 16.5 GHz, respectively. They also show that the dominant modes has the phase delays of about $\pi$ and $3\pi$ at 6.9 GHz and 16.5 GHz, respectively.

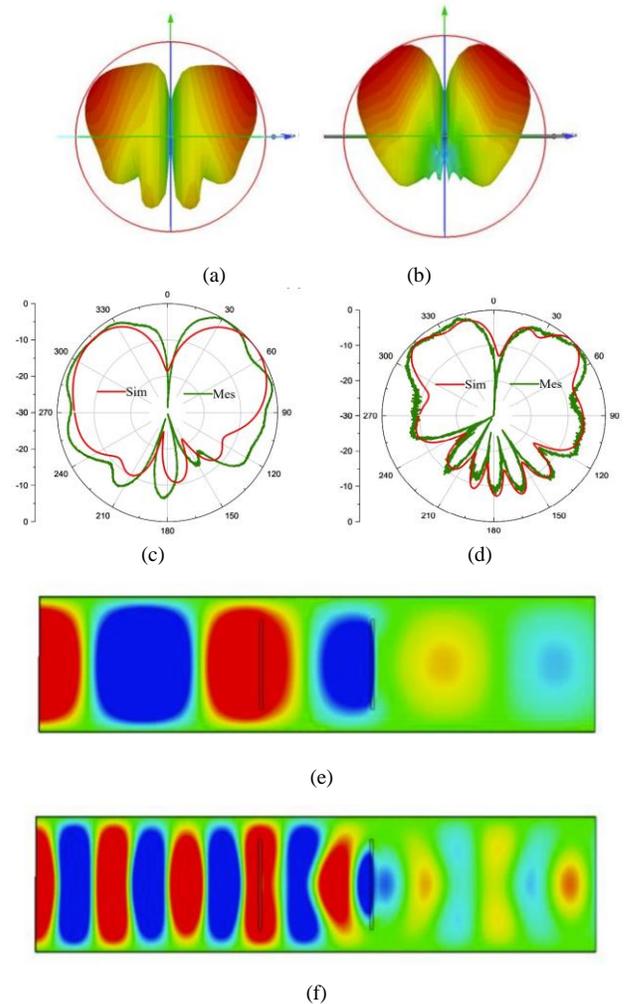

Fig. 2. Radiation patterns and electric fields of the two-slot DFW/SIW: (a) Simulated DFW pattern at 6.9 GHz, (b) Simulated DFW pattern at 16.5 GHz, (c) Simulated and measured SIW E-plane pattern at 6.9 GHz and (d) Simulated and measured SIW E-plane pattern at 16.5 GHz, (e) E-field at 6.9 GHz and (f) E-field at 16.5 GHz.

. The calculated phase delays by (9), the showed phase delays in Fig. 2(c) and (d) are consistent. These phase delays are $\pi$ and $3\pi$ at 6.9 GHz and 16.5 GHz, respectively. The null radiation direction predicted by (10) based on these phase delays is the same as the simulated and measured null radiation. All these consistent results indicate that when operating frequencies are 6.9 GHz and 16.5 GHz, the slow-wave dominant mode in the DFW produce leaky-wave radiation.

All the above results show that slow-wave dominant mode in a non-uniform antenna can produce the leaky-wave radiation.

## IV. CAN SLOW-WAVE FUNDAMENTAL MODE EXPLAIN PATTERN CHARACTERISTICS OF PLWA

When a PLWA has slow-wave radiation, the fundamental mode is slow-wave. In classic PLWA theory, its pattern characteristics of the PLWA, such as the direction of its main lobe, is usually explained with the fast-wave space harmonic. This Section provides a proof that the difference of the phase constants between the slow-wave fundamental mode and the fast-wave space harmonics has not any effect on the radiation pattern characteristics of a PLWA.

The fundamental mode and its space harmonics share the same attenuation constant and different phase constants. The phase of the space harmonic is as follow [4]

$$\beta_n = \beta_0 + \frac{2n\pi}{p}, \quad (n = 0, \pm 1, \pm 2, \cdots) \quad (11).$$

Therefore, there are two kinds of phase delays between any two adjacent radiating elements of a PLWA: one calculated using the fundamental mode and the other using fast-wave space harmonics.

For a PLWA with $N$ elements, its array pattern can be expressed by [26]

$$F(u) = \left| \frac{\sin(0.5Nu)}{\sin(0.5u)} \right| \quad (12)$$

$$u = k_0 p \cos\theta - \beta p \quad (13)$$

where $u$ is the total phase difference of two radiating fields from adjacent two element antennas, and $F(u)$ is a periodic function with a periodicity of $2\pi$.

The phase difference based on $n$th fast-wave space harmonic is as follow

$$u_n = k_0 p \cos\theta - \beta_n p \quad (14).$$

Substituting (11) into (14), one has

$$u_n = u_0 - 2n\pi \quad (15),$$

where $u_0$ is the phase difference based on the fundamental mode. Based on the periodic characteristic of $F(u)$, one has

$$F(u_n) = F(u_0) \quad (16).$$

It means that the difference of the phase constants between the slow-wave fundamental mode and the fast-wave space harmonics has not any effect on the radiation pattern characteristics of a PLWA. On the other hand, the fundamental mode has the same attenuation constant as the space harmonics, but different phase constants. Therefore, it might be possible that the fundamental mode could fully replace the role the harmonics used to play in the classic PLWA theory.

## V. WHICH IS REASONABLE: FUNDAMENTAL MODE PRODUCE OR SPACE HARMONIC PRODUCE

Since slow-wave modes in a non-uniform structure antenna can produce radiation, and the difference of the phase constants between the slow-wave fundamental mode and the fast-wave space harmonics has not any effect on the radiation pattern characteristics of a PLWA, it may be reasonable that the fundamental mode rather than its space harmonics, produces the radiation of a PLWA, regardless of whether the fundamental mode is fast-wave or slow-wave. The reason for it is as follows.

1) The fundamental mode has a definite physical significance since it satisfies the boundary condition of a PLWA. On the other, any one space harmonic does not satisfy the boundary conditions individually, so that any individual space harmonic does not exist alone in a periodic structure [4]. Moreover, any individual space harmonic has not a field distribution corresponding to an actual antenna. It means that the space harmonics have less physical significance compared to the fundamental mode.

2) It would be impossible to tell whether any individual space harmonic radiate or not. Whether or not wave radiates depends on the continuity of the structure boundary. Any individual space harmonic is only part of the overall electromagnetic wave, so it has nothing to do with boundary continuity. Therefore, a space harmonic cannot be used to determine whether a PLWA radiates or not. If one asks what kind of modes can produce radiation, even if the answer to this question is not the fundamental mode, it cannot be a space harmonic, regardless of whether the space harmonic is fast-wave or slow-wave. On the other hand, the fundamental mode satisfies boundary conditions and has a field distribution associated with an actual antenna, and it would be evident to determine whether the fundamental mode radiates or not. Moreover, since the field distribution can also be calculated with simulation tools, the fundamental mode provides a more visual insight than space harmonics.

3) In principle, all space harmonics in a PLWA share the same importance. If two spatial harmonics are different types of wave, saying one fast-wave and the other slow-wave, the radiation of the PLWA cannot be produced/determined by only one and not the other. In addition, since a space harmonic is only part of the entire electromagnetic wave, it is not easy to understand that only several space harmonics can characterize the radiation characteristics of the entire electromagnetic wave. The fundamental mode, however, can fully characterize the entire electromagnetic wave in a PLWA, including the radiation.

Since the fundamental mode rather than any space harmonics has a definite physical significance, it is the fundamental mode,

not its space harmonics, that produces the radiation in a PLWA even if the fundamental mode is a slow-wave.

## VI. GENERALIZED FUNDAMENTAL MODES

Besides the minimum periodicity, any one periodic function has any number of other periodicity. Can the fundamental mode have multiple phase constants?

The field in a periodic structure can be rewritten in the following form

$$\mathbf{E}(x,y,z) = \mathbf{E}_0(x,y,z)e^{-j\beta_0 z} = \mathbf{E}_0(x,y,z)e^{j\frac{2n\pi}{p}z}e^{-j(\beta_0+\frac{2n\pi}{p})z} \quad (17).$$

Defining a set of generalized fundamental modes as

$$\mathbf{E}_{gn}(x,y,z) = \mathbf{E}_0(x,y,z)e^{j\frac{2n\pi}{p}z} \quad (18)$$

the field in a periodic structure can be expressed with any one generalized fundamental mode multiplied by a phase factor of $e^{-j\beta_n z}$ as follow

$$\mathbf{E}(x,y,z) = \mathbf{E}_{gn}(x,y,z)e^{-j\beta_n z} \quad (19)$$

where $\beta_n$ is phase constant of the generalized fundamental modes, and is just the same as the phase constant of space harmonics. When $n=0$, $\mathbf{E}_{g0}(x,y,z) = \mathbf{E}_0(x,y,z)$. Based on (2) and (18), one can find

$$\mathbf{E}_{gn}(x,y,z+p) = \mathbf{E}_{gn}(x,y,z) \quad (20).$$

Therefore, $\mathbf{E}_{gn}(x,y,z)$ is still a periodic function with a periodicity $p$.

Since the formula (18) shows that the generalized fundamental modes have the same magnitude as the fundamental mode, and the two kinds of modes differ by only one phase factor, the generalized fundamental modes still satisfy the same boundary condition as the fundamental mode, so the generalized fundamental mode have physical significance as the fundamental mode.

The phase constants of the generalized fundamental modes are consistent with the phase constants of space harmonics. Accordingly, if wave phenomenon related to phase characteristic in a periodic structure is explained well with a space harmonic, the wave phenomenon would be also explained well with a generalized fundamental mode. Therefore, the generalized fundamental modes could replace the roles that were once played by the space harmonics, and it would be much reasonable than space harmonics due to its physical significance.

The introduction of the generalized fundamental modes is based only on the periodicity of the fundamental mode, and is independent of the space harmonics. On the other hand, the generalized fundamental modes also provide a basis for the usability of the space harmonic explanation of radiation. With the help of the generalized fundamental modes, it can be showed why the logic of the space harmonic explanation of radiation is unreasonable but the explanation can be available.

## VII. CONCLUSION

In this study, we first prove that there is not an eigen-equation like Pythagorean theorem for the fundamental mode in a PLWA, and design an non-uniform structure antenna to show that slow-wave mode in the antenna can produce leaky-wave radiation. These results suggest an idea that the slow-wave fundamental mode in a PLWA may also produce slow-wave radiation. Secondly, it is proved that the difference of the phase constants between the slow-wave fundamental mode and the fast-wave space harmonics has not any effect on the radiation pattern characteristics of a PLWA. Thirdly, it is clarified that the fundamental mode has a more definite physical significance than its space harmonics. Accordingly, it is much more reasonable to consider the radiation in a PLWA to be produced with the fundamental mode than its space harmonics, regardless of whether the fundamental mode is fast-wave or slow-wave.

Moreover, the field in a periodic structure could be expressed as a type of generalized fundamental mode multiplied by a phase factor, which have phase constants as the same as the space harmonics. A generalized fundamental mode has physical significance as the conventional fundamental mode, and could replace the roles that were once played by the space harmonics.

The concept of the generalized fundamental mode in one dimensional periodic structures can be extended to two or three dimensional periodic structures, and the extensions are direct and easy. Moreover, because of a common mathematical base, the concept of the generalized fundamental mode could be also applied to periodic structures in solid state physics or other fields.

## ACKNOWLEDGMENT

The authors would like to thank the technical staff of the Poly-Grames Research Center at École Polytechnique de Montreal for their collaboration and support of the fabrications and measurements related to this work.